\documentclass[12pt]{article}
\usepackage{mypreamble} 
\usepackage[utf8]{inputenc}
\usepackage[style=numeric,backend=bibtex]{biblatex}
\usepackage[toc,page]{appendix}

\title{A framework for interpretation and testing of sparse canonical correlations}

\author{Nuria Senar}
\author{Mark van de Wiel}
\author{Aeilko Zwinderman}
\author{Michel Hof}

\affil{\small Department of Epidemiology \& Data Science, Amsterdam School of Public Health, Amsterdam UMC, Amsterdam, The Netherlands}

\date{October 2023}

\begin{document}

\maketitle

\abstract{
In clinical and biomedical research, multiple high-dimensional datasets are nowadays routinely collected from omics and imaging devices.  Multivariate methods, such as Canonical Correlation Analysis (CCA), integrate two (or more) datasets to discover and understand underlying biological mechanisms. For an explorative method like CCA, interpretation is key. 
We present a sparse CCA method based on soft-thresholding that produces near-orthogonal components, allows for browsing over various sparsity levels, and permutation-based hypothesis testing. Our soft-thresholding approach avoids tuning of a penalty parameter. Such tuning is computationally burdensome and may render unintelligible results. In addition, unlike alternative approaches, our method is less dependent on the initialisation. We examined the performance of our approach with simulations and illustrated its use on real cancer genomics data from drug sensitivity screens. Moreover, we compared its performance to Penalised Matrix Analysis (PMA), which is a popular alternative of sparse CCA with a focus on yielding interpretable results. Compared to PMA, our method offers improved interpretability of  the results, while not compromising, or even improving, signal discovery. he software and simulation framework are available at \url{https://github.com/nuria-sv/toscca}.
}


\section{Introduction} \label{sec:intro}

New technologies in clinical and biomedical research facilitated the collection of high-throughput omics data using DNA sequencing, RNA microarrays, or mass spectroscopy. These methods typically result in hundreds or thousands of variables per patient, yet, sample sizes remain low. As a result, the number of variables largely exceeds the number of observations.
Genomics studies concerned with finding common structures between multiple pheno- or genotypical measures call for statistical models capable of dealing with high-dimensions.

Integrative approaches such as Canonical Correlation Analysis (CCA) \parencite{originalCCA_hotelling} can contribute to improvements in diagnostics and understanding of biological mechanisms, by exploring connecting attributes between datasets. This includes genomics as well as neurological data which recently has been at the centre of many CCA applications linking brain connectivity data to genetic, demographics, behavioural or thought patterns \parencite{neurodiagnosisGenoFenoClinical_du_2020, needleCCA_wang_2020}. CCA is a multivariate method in high-dimensional analysis for exploring underlying signals relating two (or more) datasets through pairs of weight vectors. It is an immediate extension of PCA for more than one dataset, and a scale invariant adaptation of PLS. 
CCA searches for linear combinations of the data, called latent variables, that are maximally correlated to each other. 
The original variables are summarised into these lower-dimensional variables. This process may be repeated to render multiple latent variables. Methods combining dimension reduction and correlation maximisation have competitive accuracy in predicting complex traits than other conventional ML methods \parencite{sccaComparison_rodosthenus_2020}, which means that CCA is an interesting tool for high-dimensional analysis.

For such, however, computation times are prone to be high and 
results may be difficult to extrapolate, generalise or test. Furthermore, interpreting the estimated weights is far from trivial in large dimension problems. This problem persists in CCA applications as there may be many possible linear combinations of variables maximising correlations. Hence, the probability of selecting highly correlated noise increases and so does the in-sample canonical correlation estimate. In these scenarios, the presence of redundant variables is dealt with through sparse constraints dealing with regularisation, variable selection or a combination of both. 

Recent methods for CCA search for complex associations, i.e. nonlinear or supervised. 
However, these generally are concerned with prediction and show lack interpretability, or assume some type of structure or classification. As we are concerned with finding interpretable results from an exploratory analysis, we focus on penalised alternatives which render sparse weights for both datasets.

Sparse extensions to CCA include lasso or elastic net penalisation methods for parameter shrinkage and variable selection \parencite{waaijenborg2007penalized, scca_parkhomenko, pma_witten}. The sparsity is chosen through cross-validation techniques of the penalty parameters. 
Said techniques are known to be unstable both in terms of the estimation of the penalty parameters \parencite{estSparsity_vanNee_2022} and that of the coefficients \parencite{sparsityVsStability_xu}. Not only does this affect interpretability of the results, it also means that the permutation testing framework is ill-behaved. 

We address these concerns by imposing a threshold on the support of the canonical vectors using soft-thresholding, rather than using a penalty parameter. Hence, we introduce sparsity into our canonical vectors by stating the number of nonzero weights, keeping the number of selected variables equal through permutations and promoting interpretable results via direct control over the number of selected variables. In addition, to achieve Type-I error control, we also show that using the out-sample correlation, instead of the in-sample correlation, accounts for spurious associations. We propose a fast estimation scheme based on the NIPALS \parencite{nipals_wold} algorithm, essential for efficient testing in high-dimensional CCA.

With simulations, we evaluated signal recovery and Type-I error control and compared its performance to the popular Penalised Matrix Analysis (PMA) method \parencite{pma_witten}. Moreover, we applied our algorithm to real data on gene expression and drug sensitivity measures to study the performance of our sparse CCA. We found that, compared to PMA, fixing the number of nonzeros improved stability of the shape and size of the canonical weights for increasingly large matrices.

\section{Methods} \label{sec:method}

Suppose we have two data matrices $\bX_1 \in \mathbb{R}^{n\times p}$ and $\bX_2 \in \mathbb{R}^{n\times q}$ containing respectively $p$ and $q$ variables from $n$ samples from which we want to extract a sequence of $K$ pairs of canonical vectors $\left\{( \balpha_{1}, \bbeta_{1}), \dots, ( \balpha_{K},  \bbeta_{K})\right\}$, where $K \leq \text{min}(p, q)$. 
The $k^{th}$ pair of canonical variables is given by $\bgamma_{k} = \bX_1\balpha_{k}$ and $\bzeta_{k} = \bX_2\bbeta_{k}$. The correlation between this pair of canonical variables, referred to as canonical correlation, is given by

\begin{equation}\label{eq:CCA}
\begin{split}
    \rho_k = \frac{\balpha_k^T\bX_1^T\bX_2\bbeta_k }{\sqrt{\balpha_k^T\bX_1^T\bX_1\balpha_k }\sqrt{\bbeta_k^T\bX_2^T\bX_2\bbeta_k }}    
\end{split}
\end{equation}

\noindent
The goal of CCA is to choose the weights $\mathbf A=(\balpha_{1}, \ldots,\balpha_{k}, \ldots,\balpha_{K})$ and $\mathbf B =(\bbeta_{1}, \ldots,\bbeta_{k},\ldots, \bbeta_{K})$ such that the correlation between all canonical vectors is maximised under the restriction that the columns in the sets $\left(\bgamma_{1}, \ldots, \bgamma_K\right)$ and $\left(\bzeta_1. \ldots, \bzeta_K\right)$ are orthogonal. Generally, the weights are estimated such that the first pair of canonical vectors has the highest canonical correlation and with each succeeding pair the canonical correlation decreases.


\subsection{Nonlinear Iterative Partial Least Squares and CCA}

In this paper, we consider the CCA problem in a regression framework in which pairs of
canonical vectors are sequentially estimated with an alternating regression procedure. This technique, known as Nonlinear Iterative Partial Least Squares (NIPALS) \parencite{nipals_wold}, starts by initialising one of the canonical vectors, $\balpha^{(0)}$, and computing $\bbeta$ given $\balpha^{(0)}$. 
The estimation of the weights $\bbeta$ is equivalent to a simple least square problem. To obtain the first pair of canonical vectors, we use the equivalence between maximising equation \eqref{eq:CCA} and the optimisation problem 


\begin{align} \label{eq:paper5_E1}
    (\hat{\balpha}, \hat{ \bbeta}) = \arg\min_{ \balpha,  \bbeta} \sum_{i=1}^n \bigl( {\mathbf{x}}_{1,i}  \alpha_i -      {\mathbf{x}}_{2,i}  \beta_i   \bigr)^2,
\end{align}

\noindent where the canonical variables are required to have unit norm to have the same constraints as in equation \eqref{eq:CCA}.
In the alternating regression procedure, we initialise and fix vector \(\balpha^{(0)}\) and scale \( \bgamma\) to have unit norm after step \ref{st:gamma0} in algorithm \ref{alg:scca}, i.e. 

\[ \bgamma^{(0)} = \frac{\bX_1\balpha^{(0)}}{\sqrt{\balpha^{(0)T}\bX_1^T\bX_1\balpha^{(0)}}}\]

\noindent By fixing $\alpha^{(0)}$, equation \eqref{eq:paper5_E1} reduces to a simple (least squares) regression problem \parencite{sparse_wilms_2015}. An estimate of $\bbeta$ obtained as

\begin{equation}\label{eq:paper5_eq1}
\bbeta^{(1)} = \left(\bX_2^T\bX_2\right)^{-1}\bX_2^T \mathbf \bgamma^{(0)}
\end{equation}


\noindent Vice versa, we obtain $\bzeta^{(1)}$ and $\balpha^{(1)}$ fixing $\bbeta^{(1)}$. The estimated vectors describe the strength of linear association between the matrices.
This process is then repeated until convergence of some tolerance measure.

Generalised to $k>1$, we initialise \(\balpha_k\) as \(\balpha_k^{(0)}\) to then repeatedly fix and re-estimate new weights to obtain a sequence
\(\{(\balpha_k^{(0)}, \bbeta_k^{(0)}), (\balpha_k^{(1)}, \bbeta_k^{(1)}), \ldots,\}\) that is monotonically convergent \parencite{nipals_convergence_hanafi_2007} for each component. The first canonical vector of $\bX_1$, $\balpha_k^{(0)}$, can be initialised randomly, with uniform weights or with some type of matrix decomposition.

Many penalised alternatives based on lasso \parencite{scca_parkhomenko, pma_witten} or elastic net \parencite{waaijenborg2007penalized} have been proposed to deal with high-dimensional CCA. Both approaches impose sparsity in the weights $\balpha_k$ and $\bbeta_k$ by penalising the model used in equation 3. To obtain a certain sparsity, it is therefore necessary to search for the corresponding penalty parameter.  As an alternative, we propose to introduce a soft-thresholding penalty to the regression formula (\ref{eq:paper5_eq1}). This penalisation allows us direct control on the number of nonzero weights in both $\balpha_k$ and $\bbeta_k$. Not only will this improve the interpretation of the results, it also speeds up NIPALS algorithm since we do not have to search for the penalty that corresponds to a particular number of nonzero weights. Additionally, this allows us to use permutations for hypothesis testing. 

\begin{algorithm}[H]
    \caption{Thresholded Ordered Sparse CCA (TOSCCA)}\label{alg:scca}
    \textbf{Input. } $\bX_{1,s}$, $\bX_{2,s}$, $\balpha^{(0)}$, $p_{\balpha}$, and $q_{\bbeta}$\\
    \textbf{Output. } $\balpha_k^*$ and $\bbeta_k^*$ \\
    $t \gets 1$, $\theta << 1$, $\varepsilon = 10^6$, $\rho^{(0)}0 \gets 0$
    \begin{algorithmic}[1]
    \While{$\varepsilon > \theta$} \Comment{Changes larger than tolerance measure}
    \State $\bgamma \gets \bX_{1,s}\balpha^{(t-1)}$ \label{st:gamma0}
    \State $\tilde{\bbeta}^{(t)}\gets \bX_{2,s}^T\bgamma$ \label{step:beta_gamma}
    \State $\bbeta_k^{(t)} \gets \mathbbm{1}_{|\tilde{\bbeta}^{(t)}|>q_{\bbeta}}\tilde{\bbeta}^{(t)}  - q_{\bbeta}$ \label{st:soft_beta}
    \State $\bzeta_k \gets \bX_{2,s}\bbeta_k^{(t)}$ \Comment{Standardise canonical variable for $\bX_2$} \label{st:zeta}
    \State $\tilde{\balpha}^{(t)} \gets \bX_1^T\bzeta_k$ \label{setep:alpha_zeta}
    \State $\balpha_k^{(t)} \gets \mathbbm{1}_{|\tilde{\balpha}^{(t)}|>p_{\balpha_i}}\tilde{\balpha}^{(t)} - p_{\balpha}$ \label{st:soft_alpha}
    \State $\bgamma_k \gets \bX_{1,s}\balpha_k^{(t)}$ \Comment{Standardise canonical variable for $\bX_1$}
    \State $\rho^{(t)} \gets cor(\bgamma_k, \bzeta_k)$
    \State $\varepsilon \gets \rho^{(t)} - \rho^{(t-1)}$ 
    \State $t = t+1$
\EndWhile\label{sccaWhile}
\State \textbf{return} $(\balpha_k^*, \bbeta_k^*)$\Comment{The canonical vectors}
\end{algorithmic}
\end{algorithm}
\noindent To add the soft-threshold penalty, we ignore the collinearity of our data by assuming that $\bX_1^T\bX_1 = \bI_p$. As with $\bX_2$, simplifying the regression from equation \eqref{eq:paper5_eq1} into steps \ref{step:beta_gamma} and \ref{setep:alpha_zeta} \parencite{ignoreCollinearity_dudoit}. We calculate the optimal weight coefficients in algorithm \ref{alg:scca}, through a modified NIPALS with soft-thresholding (steps \ref{st:soft_alpha} and \ref{st:soft_beta}) based on threshold parameters $p_\alpha\in \left\{1,2,\dots, p\right\}$ and $q_{\beta}\in \left\{1,2,\dots, q\right\}$. We can show that the relationship between the in-sample canonical correlation and progressively larger $p_{\alpha}$, given $q_{\beta}$, is nonconcave and increases for non-sparse solutions. Therefore, we use the out-sample canonical correlation which indeed shows a convex trajectory, implying decrease of the canonical correlation, as more irrelevant variables are included.

Through this algorithm, smaller choices of $\left(p_{\alpha}, q_{\beta}\right)$ yield canonical vectors which are subsets of dense alternatives when there is a signal, keeping one penalty fixed. That is, for threshold choices $p_{\alpha,1} \leq \dots \leq p$ and some fixed $q_{\beta}$, both in the simulation study and the real application, $supp(\balpha(p_{\alpha,i})) \subseteq supp(\alpha(p_{\balpha,j}))$ if $i \leq j$. This property is useful for interpretation of the results, as it shows selection stability of the larger contributors.

\subsection{Estimating multiple canonical variates}\label{sec:method_sparsity}

In high-dimensional settings, finding the \textit{true} canonical weights linking both datasets is particularly difficult as, in practice, there are many possible competitive combinations, rendering comparable canonical correlations. Furthermore, we wish to balance the percentage of explained variance to the number of selected variables, ruling out tuning of a penalty parameter. Instead, we choose soft-thresholding, which allows the user to search over and compare results from multiple specific sparsity levels at the same time. 

From the computational perspective, the NIPALS algorithm allows the simultaneous and efficient estimation of the canonical vectors for different combinations of penalties by defining vector pair $(\mathbf{p}_{\alpha}, \mathbf{q}_{\beta})$. Then the canonical vectors for each component become matrices ($\bA_k, \bB_k$) for which each column represents a $(p_{\alpha,i}, q_{\beta,i})$ pairing. In a single run of the NIPALS algorithm, it is possible to estimate the canonical vectors for several sparsity levels. That is, steps \ref{step:beta_gamma} and \ref{setep:alpha_zeta} in algorithm \ref{alg:scca} become $\bB = \bX_2^T\mathbf{\Gamma}$ and $\bA = \bX_1^T\mathbf{Z}$, where $\mathbf{\Gamma}$ and $\mathbf{Z}$ are matrices of latent variables from canonical weights with different sparsity levels.


We calculate the canonical weights and correlations for later components ($k \geq 2$) deflating the data to account for the variance explained in previously estimated latent variables. 
We deflate the matrices as

\begin{equation}\label{eq:res_reg}
    \bX^{(k+1)}_1 =(I_p - \bgamma^{(k)}(\bgamma^{(k)T}\bgamma^{(k)})^{-1}\bgamma^{(k)T})\bX^{(k)}_1,
\end{equation}

\noindent where $\bX_{_1}^{(1)} = \bX_1$. Matrix $\bX_2$ is deflated following the same scheme. 

In the original unpenalised version of NIPALS, this deflation would make latent variables orthogonal for different components. However, it is well known that introducing sparsity compromises this property \parencite{sparseOrth_Jolliffe} as it is the case with other sparse CCA methods. 
Consequently, the standard measure of cumulative percentage of explained variance (CPEV)\footnote{Where CPEV is defined as $tr(\bgamma_{1:k}^T\bgamma_{1:k})/tr(\bX_1^T\bX_1)$, and equally for $\bzeta$ and $\bX_2$.}, which is used to determine the number of components \parencite{sPCALowRank_shen}, may be inaccurate due to the presence of repeated information. That is, as orthogonality of the canonical vectors can no longer be guaranteed, new components do not necessarily contain new information, and hence latent variables may be correlated. We propose the following alternative measure to adjust for repeated information for $k = 2, \dots, K$:

\begin{align}\label{eq:adj_cpev}
    &\text{CPEV}_{\text{adj}}(\bgamma_k) = CPEV(\bgamma_k)\cdot\prod_{i<k}(1 - |cor(\bgamma_i, \bgamma_k)|)
\end{align}
\vspace{-2em}

\subsection{Permutation testing}\label{sec:perm}

To assess the estimated correlations, we test the null hypothesis of no correlation between the datasets, and their deflated counterpart for subsequent components via permutation testing. We permute one of the datasets and re-estimate the canonical correlation to approximate the distribution of the correlations under the null. Since the canonical correlation estimate is affected by the number of nonzero weights, using the number of nonzeros as the original analysis makes the permuted correlations comparable between themselves and the original estimate. 

Standard penalties, such as the lasso or the elastic net, optimise the combination of weights and variable selection to match the corresponding dataset. This yields null distributions which are contingent on sparsity levels and, thus may lead to incorrect assessment of the estimated canonical correlations, as the same penalty across permutation may not return the same sparsity level\footnote{See Figure A in the supplementary material.}.

The canonical correlation estimate is non-decreasing as a function of variables selected. Controlling over the number of the selected variables avoids catering the dimension reduction to the indiosyncracies of the data. We argue that setting a more direct penalty over variable selection together with an appropriate residualisation scheme has several advantages. Mainly these are improvements in subsequent signal detection, interpretability and assessment of the relationships found without interference form. This scheme returns appropriate type I error rates. 

Multi-component canonical correlation analysis requires testing for each component. 
Due to the high-dimensional data we expect the gaps between quantiles of the null distributions to be small; under the null distribution, the estimated correlations will have similar values. We have empirical evidence supporting this statement coming from the permutation distributions amongst different correlation estimates looking very similar. Hence we determine that a simple Bonferroni correction will suffice to manage multiple testing concerns. We address multiple testing concerns using the statistic for the largest correlation as threshold for the rest.

\section{Simulations} \label{sec:simulations}

\begin{figure}[!t]
    \centering
    \includegraphics[width=0.9\textwidth]{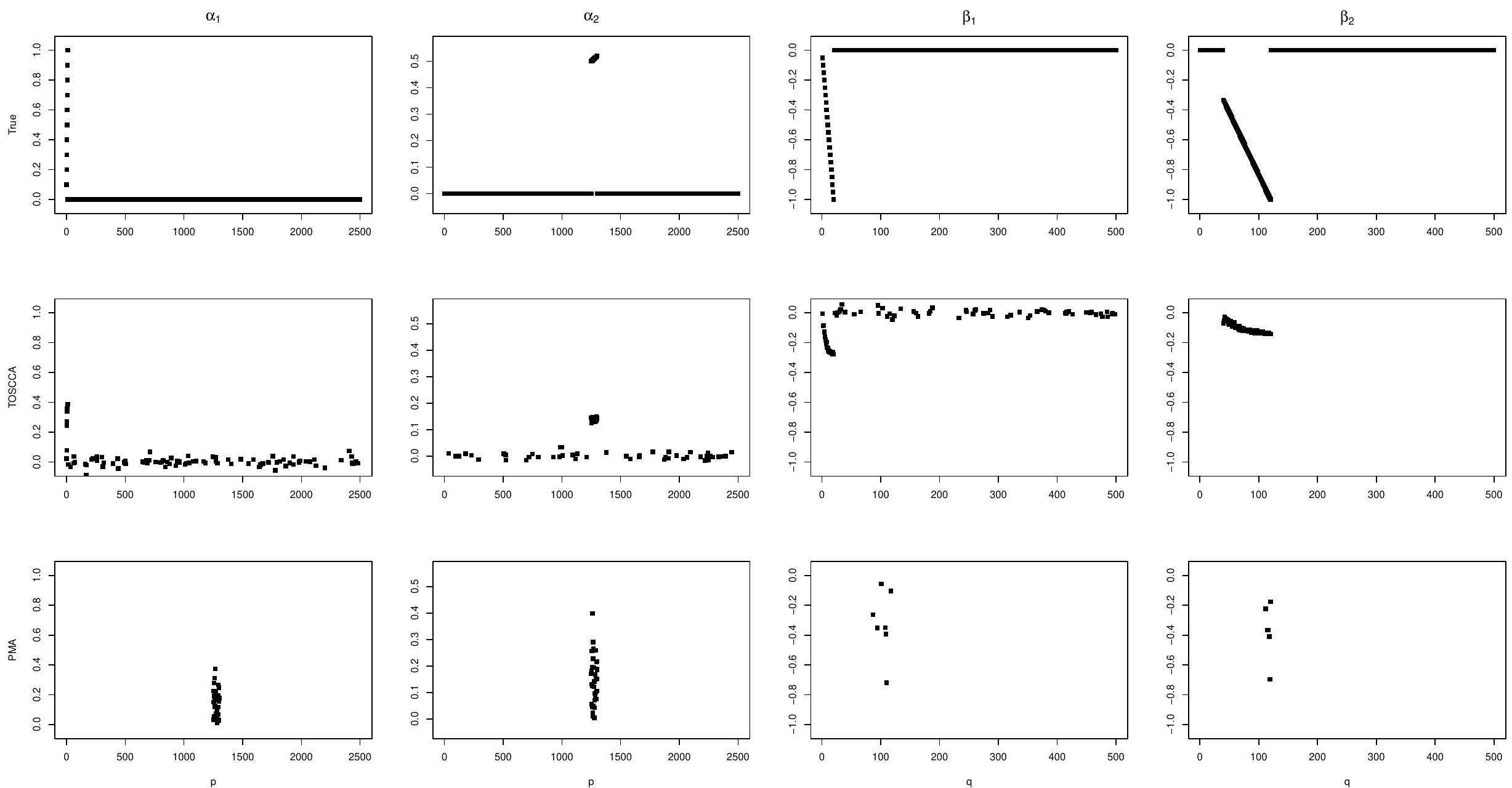}
    \caption{True (top) and estimated canonical vectors for TOSCCA (middle) and PMA (bottom).}
    \label{fig:true_estim}
\end{figure}

We simulated three true components of different sizes for data with $n=100$, $p=2500$ and $q=500$. We analysed the simulated data using the approach from section \ref{sec:method}, from now on referred to as TOSCCA for Thresholded Ordered Sparse CCA, and compared its performance to the existing method, PMA, a popular sparse CCA method used in the study of high-dimensional data aiming to improve interpretability. We examined each model's 
accuracy (Figure \ref{fig:true_estim}), 
selection stability, 
convergence and 
adjusted CPEV, from equation \eqref{eq:adj_cpev}
. 

We fixed the sparsity level for each method as $p_{\alpha} = q_{\beta} = 100$ variables for all components and found the best penalty for PMA using their built-in function. Figure \ref{fig:true_estim} displays the true signals (top) and the estimated canonical vectors by TOSCCA (centre) and PMA (bottom). For the sake of comparability, both methods had initial values drawn from a random uniform distribution. This was a minor modification to the PMA algorithm as it was originally designed to be initialised with values from an eigen rendering canonical vectors which do not differ much from their initialisation. That is, canonical vectors are effectively predetermined from the start. Since convergence irrespective of the initialisation is a desirable property for iterative algorithms, we compared the two approaches using the same random initialisation. We observed that TOSCCA consistently selects the corresponding variables for each component; when the signal involved fewer than $\{p_{\alpha}, q_{\beta}\}$ variables for each, the remaining weights were set closer to zero. Moreover, there was no overlap between signals and the canonical weights were appropriately paired. PMA, on the other hand, selected the larger signal for both components. 

We checked TOSCCA's selection stability for running the algorithm for 8 different subsamples, one for each choice of $p_{\alpha}$ while keeping $q_{\beta}$ fixed. The signal was distinctly identified regardless the number of nonzero variables. We observed the selection stability described in the previous section\footnote{See Figure B in the supplementary material}. We observed the adjusted CPEV increase for the first three components, where the signal was located, and then \textit{plateaued} for the fourth component. The auto-correlation between canonical vectors was effectively zero, reducing equation \eqref{eq:adj_cpev} to the original formula.

\noindent We assessed the validity of the estimated canonical correlations for each component through permutation testing.\footnote{See Figure C in the supplementary material.} 
We found the first three canonical correlations to be statistically different from those find in the permuted data. The fourth estimated correlation was correctly found to be not significant.


\section{GDCS data}

We applied TOSCCA to analyse data on the Genomics of Drug Sensitivity in Cancer \parencite{cancerDrugSensitivity_garnett_2012} from the GDSC project \parencite{gdsc_data_2013} aimed at identifying molecular markers of drug response. The data is comprised of drug sensitivity measures for cell lines and  their corresponding genomic profile (gene expression, methylation profiles, mutations and copy numbers) from the Catalogue of Somatic Mutations in Cancer database. 

We were interested in quantifying the associations between gene expression and drug sensitivity ($IC_{50}$) to explore how combinations of genes may affect drug effectiveness. The data is comprised of 737 samples of 49,386 gene expression measurements and 320 $IC_{50}$ values. We fixed the sparsity of the estimated canonical correlation to be of $p_{\alpha} = 100$ variables belonging to gene expression to $q_{\beta} = 20$ from the $IC_{50}$ values. These numbers were chosen to simplify analysis, limiting the number of variables to be interpreted to what we believe is feasible and to illustrate how fixing sparsity may improve results for exploratory analysis. We  ran the same analysis using PMA, where we used their proposed method based on cross-validation to find the optimal penalty parameters. These penalties rendered $800$ gene expression variables and $7$ drugs. Finally, we repeated the analysis with TOSCCA matching PMA's optimal sparsity level. 

\begin{figure*}
     \centering
     \begin{subfigure}[b]{0.475\textwidth}
         \centering
         \includegraphics[scale = 0.3]{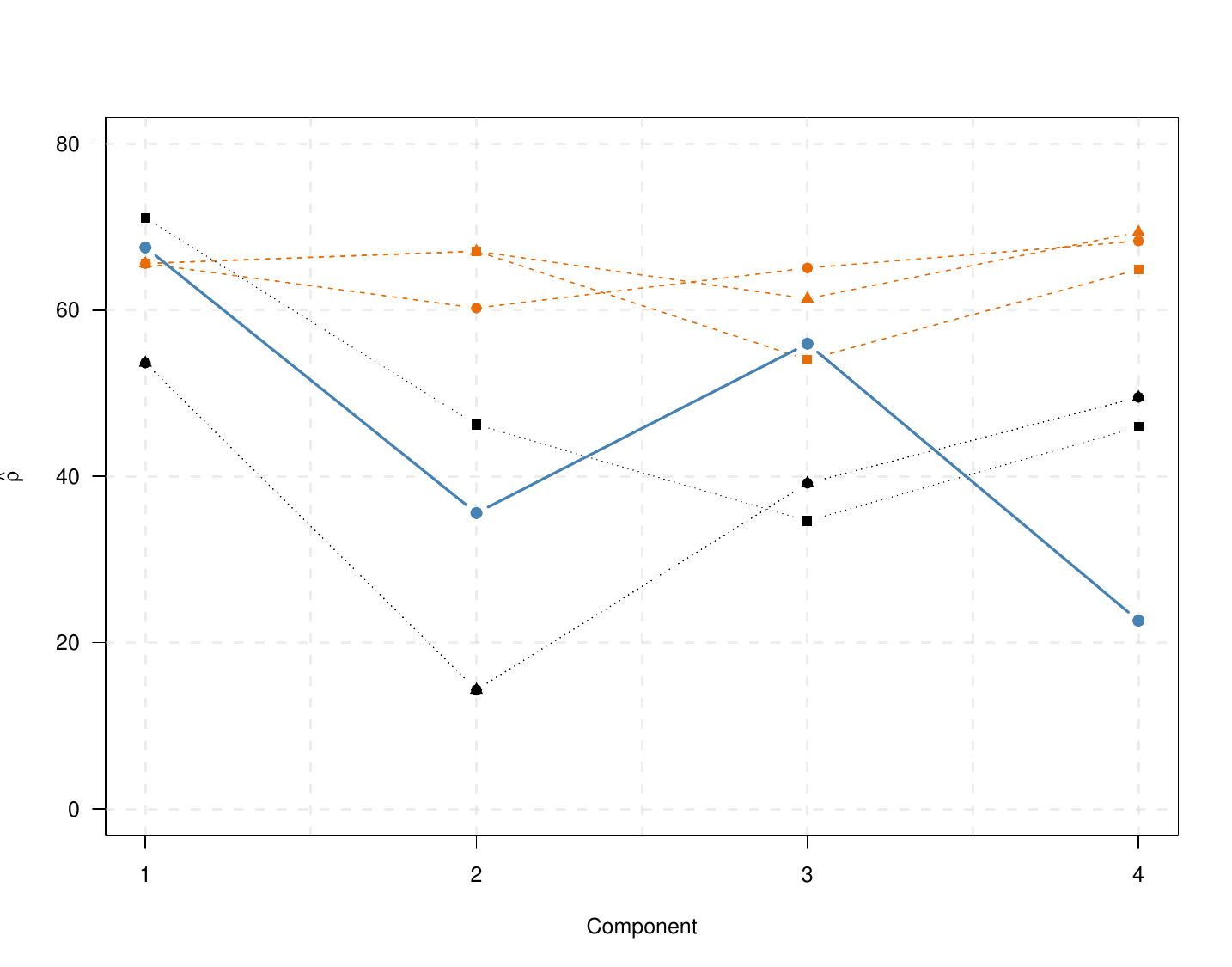}
         \caption{Out-sample CC}
         \label{fig:oos_cor_all}
     \end{subfigure}
     \hspace{.2em}
     \begin{subfigure}[b]{0.475\textwidth}
         \centering
         \includegraphics[scale = 0.3]{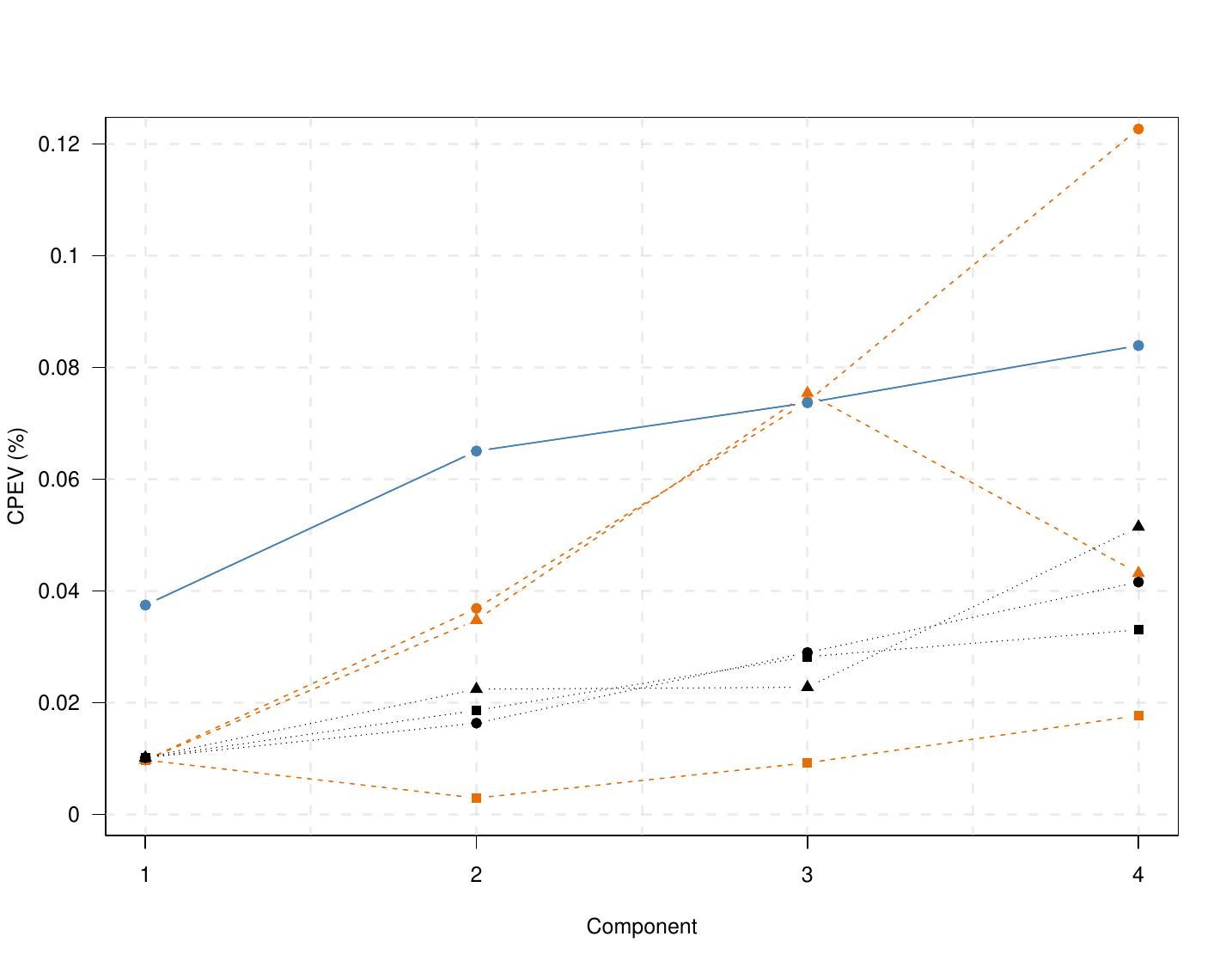}
         \caption{CPEV (\%)}
         \label{fig:cpev_drug}
     \end{subfigure}
     \hspace{.2em}
     \begin{subfigure}[b]{0.475\textwidth}
         \centering
         \includegraphics[scale = 0.3]{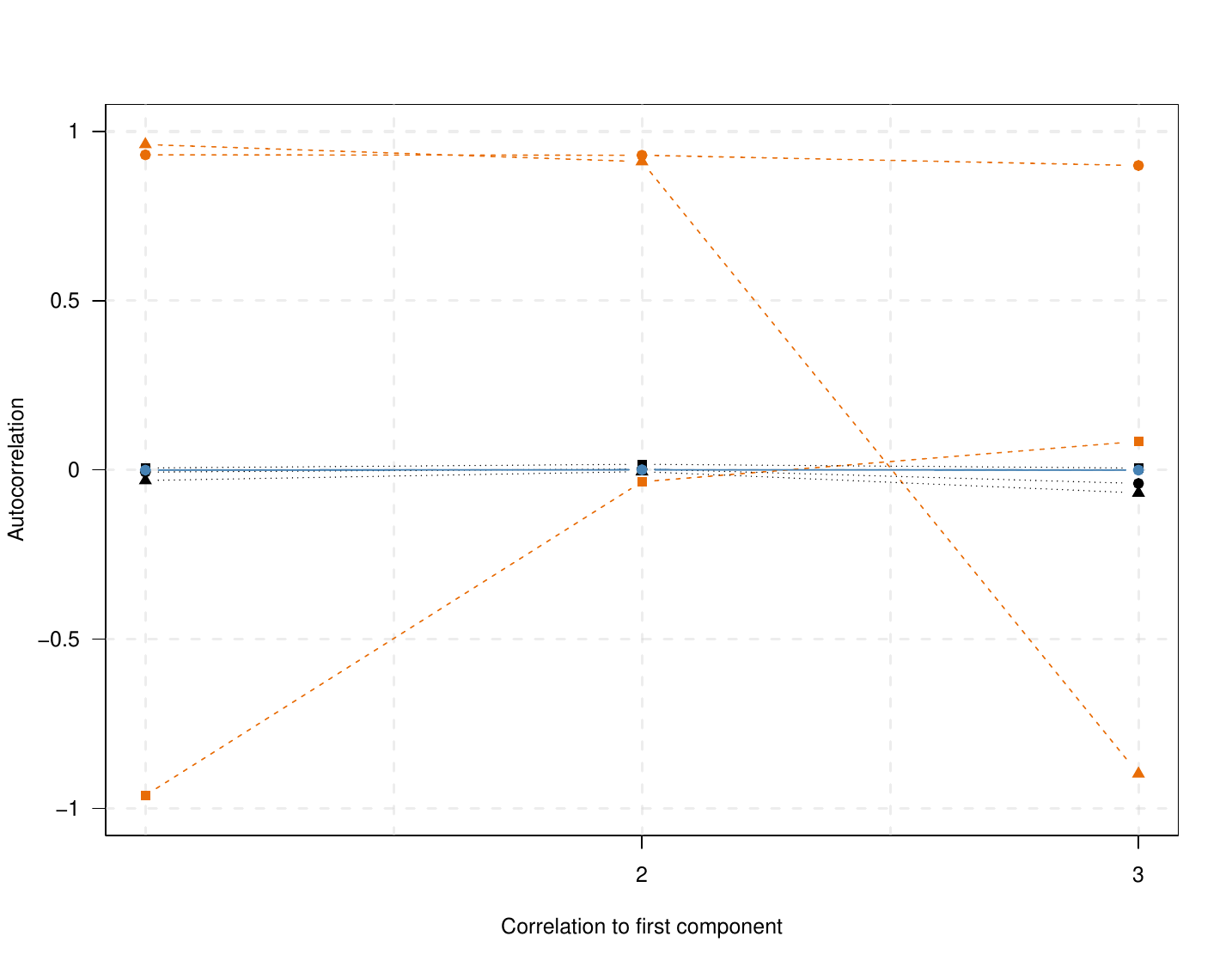}
         \caption{Correlation to $k=1$}
         \label{fig:autocorr_component}
     \end{subfigure}
     \hspace{.2em}
     \begin{subfigure}[b]{0.475\textwidth}
         \centering
         \includegraphics[scale = 0.3]{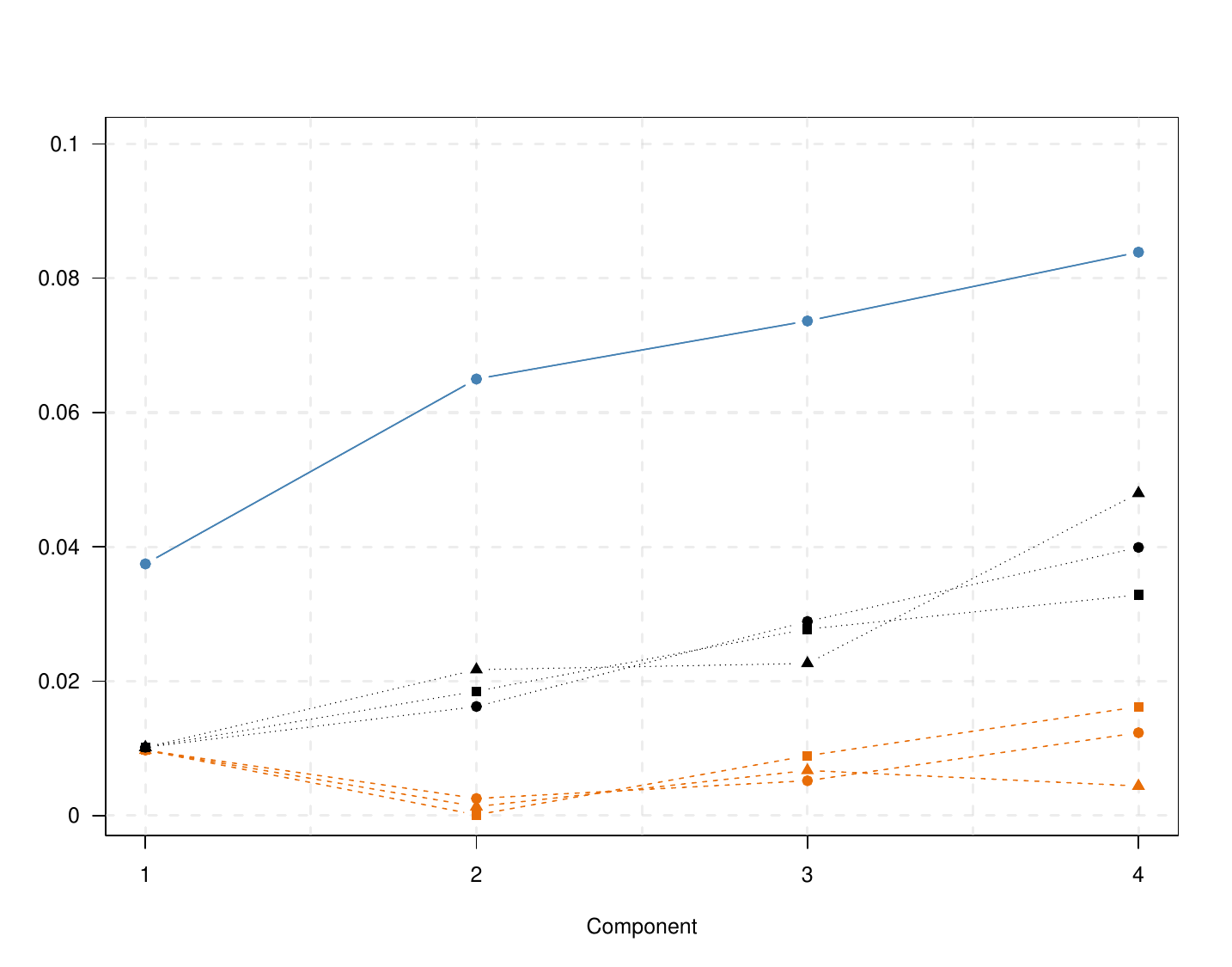}
         \caption{Adjusted CPEV (\%)}
         \label{fig:adj_cpev_drug}
     \end{subfigure}
     \begin{subfigure}[b]{1\textwidth}
         \centering
         \includegraphics[width=0.8\textwidth]{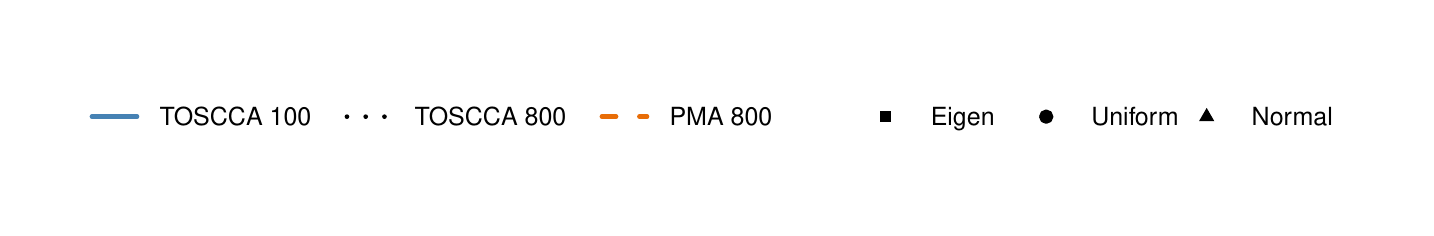}{}
     \end{subfigure}
    \caption{CCA result comparison between TOSCCA and PMA for initialisations from eigen decomposition, random uniform and random normal.}
    \label{fig:cor_figs}
\end{figure*}

We observed PMA achieve a greater correlation across components with the exception of the first component ($k=1$), 
Figure \ref{fig:oos_cor_all}. However, as previously argued, correlation alone is a weak indicator for links between high-dimensional data. The CPEV values for the four first components, Figure \ref{fig:cpev_drug}, show TOSCCA, with the default configuration (TOSCCA 100), generally outperformed all other alternatives. 
After inspection, PMA's correlation and CPEV values for subsequent components were attributed to PMA selecting virtually the same variables across components. Thus, replicating the first, usually the highest, correlation estimate.  
This is in line with what observed from PMA in section \ref{sec:simulations}, as it is prone to compute very similar components (Figure \ref{fig:true_estim}).
Figure \ref{fig:autocorr_component} shows said correlation values between the subsequent components the first one. These suggest that the variance added from subsequent components was unlikely coming from new information. 

We used the proposed adjusted CPEV in equation \eqref{eq:adj_cpev} to control for almost identical information from subsequent components. This adjusted CPEV is displayed in Figure \ref{fig:adj_cpev_drug} where the values from Figure \ref{fig:cpev_drug} account for high correlation between components, as an indicator of repeated information. After this adjustment, TOSCCA consistently outperformed PMA across the board.

We then followed with permutation testing for the estimated correlations. 
We observed all four components to be far away from the null distribution, hence deemed significant\footnote{Figure F in the supplementary material.}. Datasets of such dimensions and characteristics will most likely keep returning \textit{significant} correlations for many components as biological interactions go beyond the simple digits in this example. Nevertheless, we previously stated the advantages of keeping these links manageable in favour of interpretation. 

\begin{figure*}
    \centering
    \includegraphics[width = 0.83\textwidth]{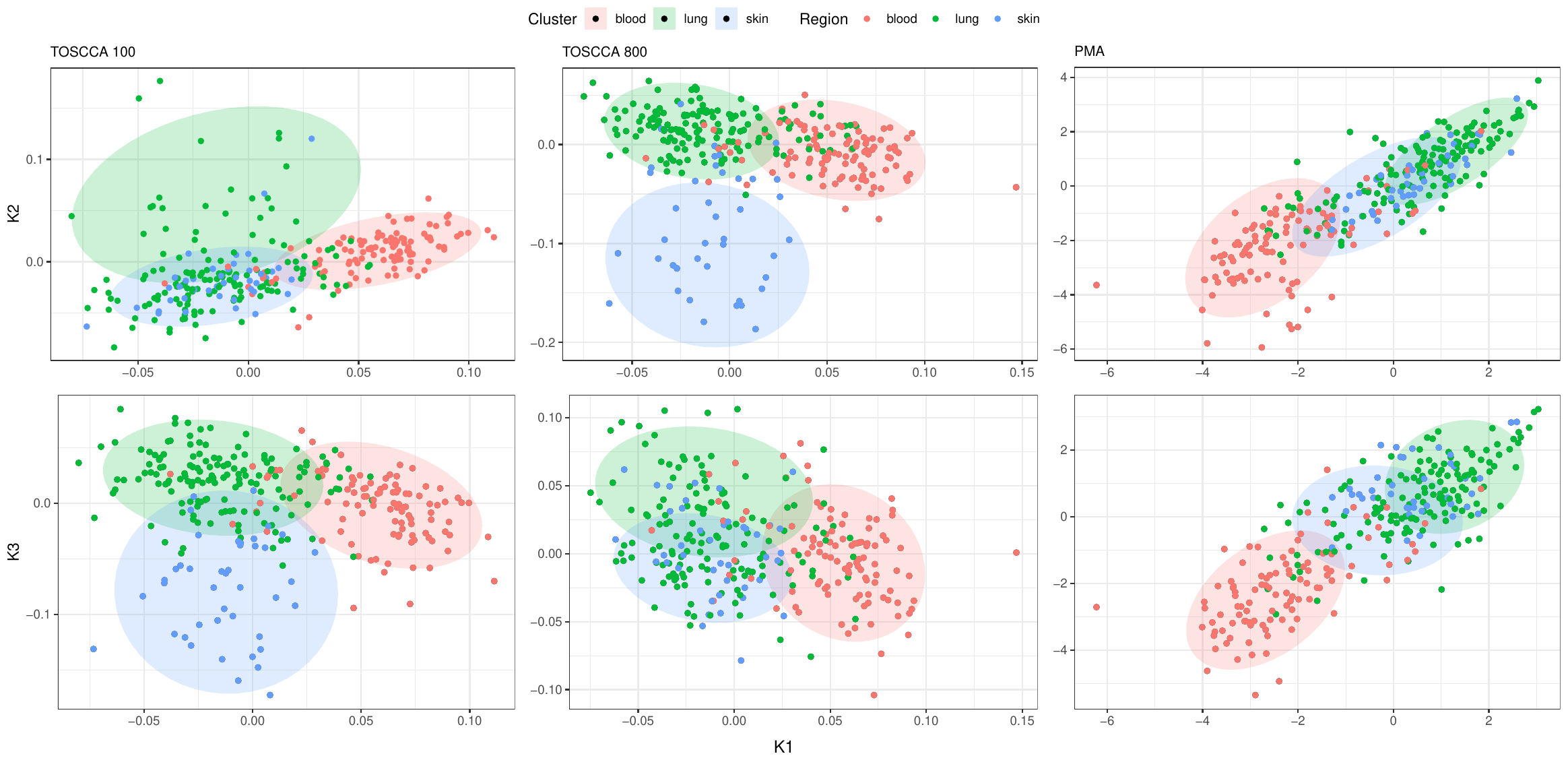}
    \caption{Latent variables plot for $k=2$ and $k=3$ against $k=1$. Sparsity levels are $p_{\alpha} = 100, q_{\beta} = 20$ (left), $p_{\alpha} = 800, q_{\beta} = 7$ (centre) and $p_{\alpha} = 800, q_{\beta} = 7$ (right).}
    \label{fig:cluster}
\end{figure*}

Last, as CCA and PCA are closely related, we used the equivalent of a score plot to observed any potential similarities represented by the estimated latent variables. We grouped observations by the region assigned to each cell line, as displayed in Figure \ref{fig:cluster}. We chose to focus on the blood, lung and skin regions as these were the ones with the most observations. Studies show idiosyncrasies in drug resistance from cancers in organ systems as they create a micro-environment which impacts drug delivery outcomes \parencite{solidCanceSensit_3_2022}, compared to that of blood cancers. 

TOSCCA found two or three different groups that were strongly associated with the cell lines' region. 
PMA (right) showed the same linearity discussed above, consequently both $k=2$ and $k=3$ look similar when plotted against $k=1$.  
These plots appear to pick up on the distinction between cancers on organ tissues and blood cancers across methods, as blood cell lines tend to remain further apart from skin and lung cell lines. Further analysis into the dynamics between gene expression and drug sensitivity measures is beyond the scope of this paper.

\section{Discussion}\label{sec:discussion}
We introduced the method TOSCCA to carry out exploratory analysis on high-dimensional data. We used the NIPALS algorithm, together with soft-thresholding to induce sparsity, for its efficiency in dealing with high-dimensional data. Our method introduces computational and interpretational attributes that ease the search and analysis of the associations integrating this data. In our method, we fix the number of nonzero canonical weights therefore promoting interpretable results and limiting the computational burden in estimation and, consequently, permutation testing. This framework allows for multiple sparsity levels to be computed simultaneously, which further facilitates the choice of sparsity. Moreover, TOSCCA shows selection stability across different choices of sparsity. 

Understanding the contribution of a variable or set of variables in penalised high-dimension analysis is complicated as different results can easily yield very similar outcomes. This is particularly true of genomic data where the correlation structure interferes with deriving inference from the results. We argue that simplifying the search is more aligned with the exploratory efforts on integrated high-dimensional data. Fixing sparsity levels achieves said goal. This approach, then, focuses on variable selection and shows selection stability both in the simulations and real data applications. 

Altogether, the above scheme supports reliable assessment of the estimated canonical correlations through permutation testing as this dimension reduction strategy has permutations be comparable and, hence, draw an appropriate null distribution. TOSCCA shows improvements in signal discovery, especially for subsequent components, and assessment when compared to the PMA method which, as established in section \ref{sec:intro}, continues to be a popular method for exploring associations in genomics datasets. 



The \textsf{R}-package \textsf{toscca} is available on github (\url{https://github.com/nuria-sv/toscca}), where we include a the script to reproduce the analysis on the simulations and the real data. 

\printbibliography

@article{ignoreCollinearity_dudoit, 
    author = "S. Dudoit and J. Fridlyand and T. P. Speed",
    title = "{Comparison of discrimination methods for the classification of
tumors using gene expression data}",
    journal = "Journal of the American Statistical Association",
    volume = "97",
    number = "457",
    pages = "77-87",
    year = "2002",
    DOI = " 10.1198/016214502753479248"
}

@article{nipals_wold, 
    author = "H. Wold",
    title = "{Estimation of Principal Components and related models by iterative least squares}",
    journal = "Journal of Multivariate Analysis",
    volume = "",
    number = "",
    pages = "391-420",
    year = "1966",
    DOI = ""
}

@Article{sparse_wilms_2015,
  author   = {Wilms, I. and Croux, C.},
  journal  = {Biometrical Journal},
  title    = {{Sparse Canonical Correlation Analysis from a predictive point of view}},
  year     = {2015},
  issn     = {03233847},
  month    = sep,
  number   = {5},
  pages    = {834--851},
  volume   = {57},
  doi      = {10.1002/bimj.201400226},
  isbn     = {3216326624},
  url      = {https://doi.org/10.1002/bimj.201400226},
}

@article{nipals_convergence_hanafi_2007, 
    author = "M. Hanafi",
    title = "{PLS Path modelling: computation of latent variables with the estimation mode B}",
    journal = "Computational Statistics",
    volume = "22",
    pages = "275-292",
    year = "2007",
    DOI = "10.1007/s00180-007-0042-3"
}

@article{sparsityVsStability_xu, 
    author = "H. Xu and C. Caramanis and S. Mannor",
    title = "{Sparse algorithms are not stable: A no-free-lunch theorem}",
    journal = "IEEE Transactions on Pattern Analysis and Machine Intelligence",
    volume = "34",
    number = "1",
    pages = "187-193",
    year = "2012",
    DOI = "10.1109/TPAMI.2011.177"
}

@article{estSparsity_vanNee_2022,
    author = {M. M. van Nee and T. van de Brug and M. A. van de Wiel},
    title = {Fast marginal likelihood estimation of penalties for Group-Adaptive Elastic Net},
    journal = {Journal of Computational and Graphical Statistics},
    volume = {0},
    number = {0},
    pages = {1-11},
    year  = {2022},
    publisher = {Taylor & Francis},
    doi = {10.1080/10618600.2022.2128809},

    URL = { 
    
        https://doi.org/10.1080/10618600.2022.2128809
    },
    eprint = { 
    
        https://doi.org/10.1080/10618600.2022.2128809
    }

}

@article{originalCCA_hotelling, 
    author = "H. Hotelling",
    title = "{Relations between two sets of variates}",
    journal = "Biometrika",
    volume = "28",
    number = "3/4",
    pages = "321-377",
    year = "1936",
    DOI = "https://doi.org/10.2307/2333955"
}

@article{sparseOrth_Jolliffe,
author = {I.T. Jolliffe},
year = {1995},
volume = {21:1},
pages = {29-35},
title = {Rotation of principal components: choice of
normalization constraints},
journal = {Journal of Applied Statistics}, 
doi = "10.1080/757584395"
}

@article{pma_witten,
    author = "D. M. Witten and  R. Tibshirani and T. Hastie",
    title = "{A penalized matrix decomposition, with applications to Sparse Principal Components and Canonical Correlation Analysis}",
    journal = "Biostatistics",
    volume = "10",
    number = "3",
    pages = "515--534",
    year = "2009",
    DOI = "https://doi.org/10.1093/biostatistics/kxp008"
}

@Article{waaijenborg2007penalized,
  author   = {Waaijenborg, S. and Zwinderman, A. H.},
  journal  = {BMC proceedings},
  title    = {{Penalized Canonical Correlation Analysis to quantify the association between gene expression and DNA markers.}},
  year     = {2007},
  issn     = {1753-6561},
  pages    = {S122},
  volume   = {1 Suppl 1},
  pmid     = {18466464},
  url      = {http://www.ncbi.nlm.nih.gov/pubmed/18466464
                  http://www.pubmedcentral.nih.gov/articlerender.fcgi?artid=PMC2367589},
}

@article{scca_parkhomenko, 
    author = "E. Pakhomenko and D. Tritchler and J. Beyene",
    title = "{Sparse Canonical Correlation Analyisis with application to genomic data integration}",
    journal = "Statistical Applications in Genetics and Molecular Biology",
    volume = "8",
    number = "1",
    year = "2009",
    DOI = "10.2202/1544-6115.1406"
}

@article{sPCALowRank_shen, 
    author = "H. Shen and J. Z. Huang",
    title = "{Sparse Principal Component Analysis via regularized low rank matrix approximation}",
    journal = "Journal of Multivariate Analysis",
    volume = "99",
    number = "",
    pages = "1115-1034",
    year = "2008",
    DOI = "10.1016/j.jmva.2007.06.007"
}

@article{cancerDrugSensitivity_garnett_2012, 
    author = "M.J. Garnett  and others",
    title = "{Systematic identification of genomic markers of drug sensitivity in cancer cells}",
    journal = "Nature",
    volume = "483",
    number = "7391",
    pages = "570-5",
    year = "2012",
    DOI = "10.1038/nature11005"
}

@article{gdsc_data_2013, 
    author = "W. Yang and others", 
    title = "{Genomics of Drug Sensitivity in Cancer (GDSC): A resource for therapeutic biomarker discovery in cancer cells}",
    journal = "Nucleic Acids Research",
    volume = "41(D1)",
    pages = "D955–D961",
    year = "2013"
}

@article{neurodiagnosisGenoFenoClinical_du_2020, 
    author = "L. Du and others", 
    title = "{Identifying diagnosis-specific genotype-phenotype associations via joint multitask sparse Canonical Correlation Analysis}",
    journal = "Bioinformatics",
    volume = "36",
    number = "",
    pages = "371-379",
    year = "2020",
    DOI = "10.1093/bioinformatics/btaa434"
}

@article{needleCCA_wang_2020,
    title = {Finding the needle in a high-dimensional haystack: Canonical Correlation Analysis for neuroscientists},
    journal = {NeuroImage},
    volume = {216},
    pages = {116745},
    year = {2020},
    issn = {1053-8119},
    doi = {https://doi.org/10.1016/j.neuroimage.2020.116745},
    url = {https://www.sciencedirect.com/science/article/pii/S1053811920302329},
    author = {H. Wang and others}
}

@article{sccaComparison_rodosthenus_2020, 
    author = "T. Rodosthenus and V. Shahrezaei and M. Evangelou",
    title = "{Integrating multi-omics data through sparse Canonical Correlation Analysis for the prediction of complex traits: A comparison study}",
    journal = "Bioinformatics",
    volume = "36",
    number = "17",
    pages = "4616-4625",
    year = "2020",
    DOI = "10.1093/nioinformatics/btaa530"
}

@article{solidCanceSensit_3_2022, 
    author = "B.J. Park and others", 
    title = "{Utilization of cancer cell line screening to elucidate the anticancer activity and biological pathways related to the ruthenium-based therapeutic BOLD-100}",
    journal = "Oncotarget",
    volume = "15",
    number = "28",
    year = "2022",
    DOI = "10.3390/cancers15010028"
}

\newpage
\appendix
\section{\\Figures}
\begin{figure}[!h]
    \centering
    \includegraphics[width = 0.8\textwidth]{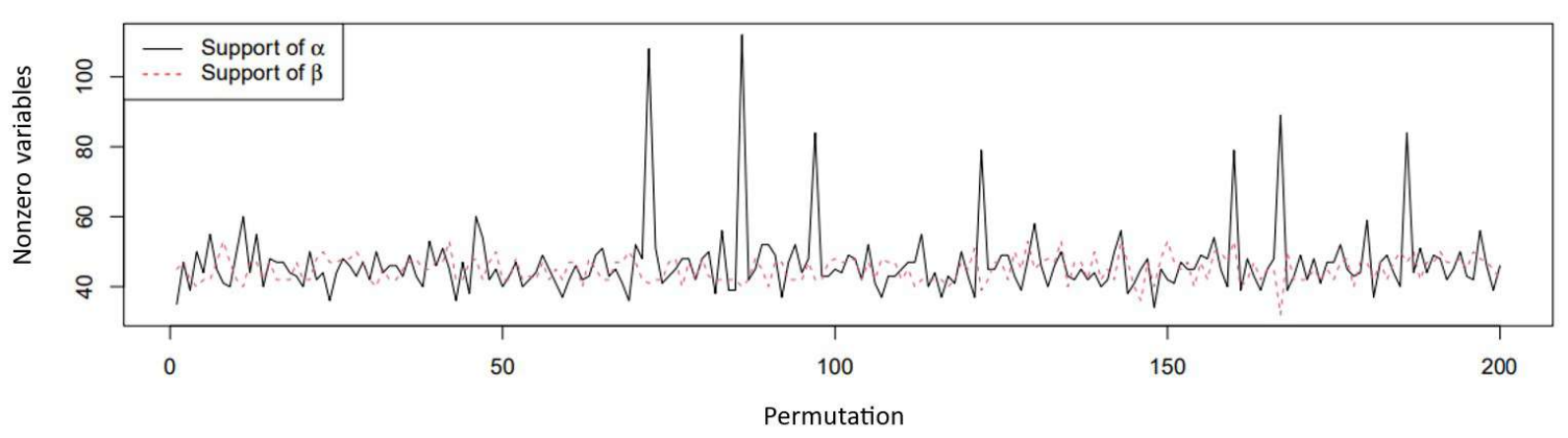}
    \captionsetup{labelformat=empty}
    \caption{Figure A: Sparsity levels for the lasso and the fused lasso penalties over different permutations.}
    \label{fig:enter-label}
\end{figure}

\begin{figure}[!h]
    \centering
    \includegraphics[width = 0.8\textwidth]{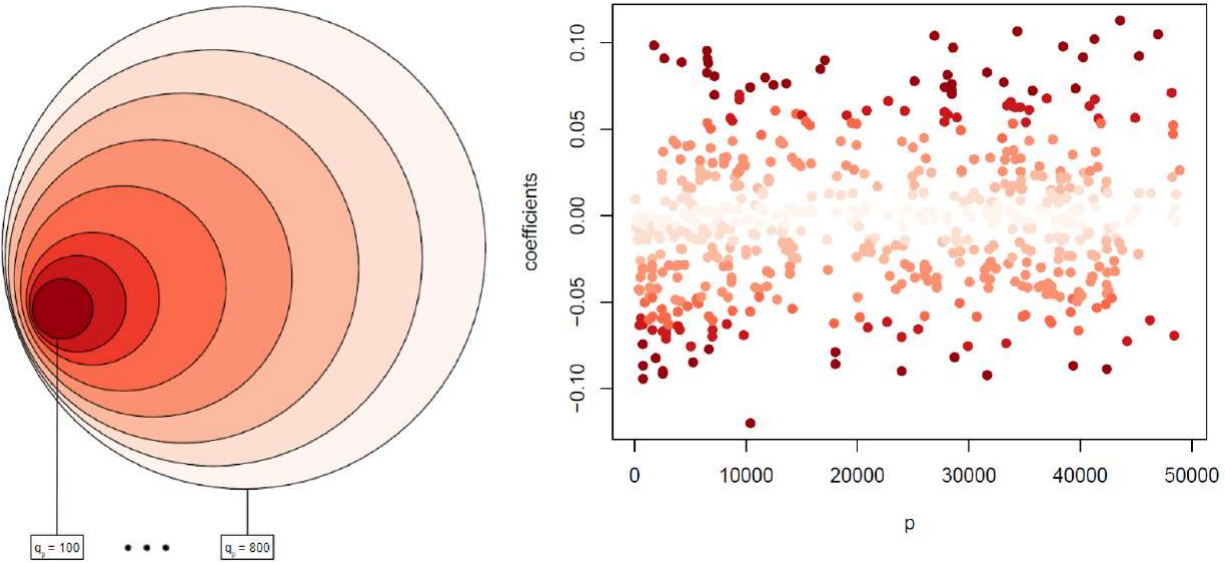}
    \captionsetup{labelformat=empty}
    \caption{Figure B: Illustration of how sparser options, such as $q_{\beta} = 100$, are subsets of denser mode ones, say $q_{\beta} = 800$. That is, all the nonzero weights in $q_{\beta} = 100$, are included in $q_{\beta} = 800$ (right). Nonzero weights for 8 different choices of the threshold parameter ($q_{\beta} = 100, 200, \dots, 800$). We see that canonical weights sparser alternatives are included in denser choices (left).}
    \label{fig:enter-label}
\end{figure}

\begin{figure}[!h]
    \centering
    \includegraphics[width = 0.8\textwidth]{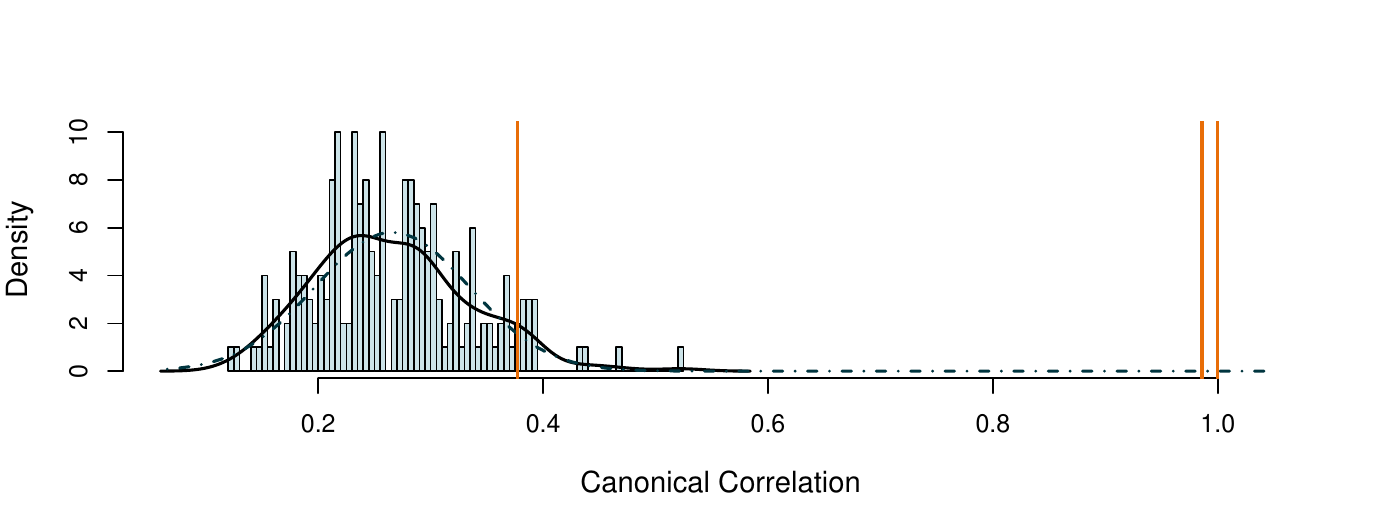}
    \captionsetup{labelformat=empty}
    \caption{Figure C: Permutations for the simulations on Section \ref{sec:simulations}. The three true components are shown to be significant while the fourth (noise) is not.}
    \label{fig:enter-label}
\end{figure}

\end{document}